\documentclass[12pt]{iopart}

\ifx\pdfoutput\undefined

\usepackage[dvips]{graphicx}

\else

\usepackage[pdftex]{graphicx}

\fi

\usepackage{setstack,iopams,amsfonts,amsthm,amssymb,times}
\usepackage{color,verbatim}
\usepackage{bbold}
\newfont{\go}{ygoth.tfm scaled 1200}  

\newtheorem{theorem}{Theorem}[section]
\newtheorem{lemma}[theorem]{Lemma}

\newcommand{\abs}[1]{\left| #1 \right|}
\newcommand{\norm}[1]{\left\|#1\right\|}

\newcommand{\DL}{\widetilde{\Delta}_L}

\newcommand{\del}[1]{\nabla_{#1}}

\begin{document}

\title{Semi-classical stability of AdS NUT instantons}
\author{Claude Warnick}
\address{DAMTP, Centre for Mathematical Sciences, Cambridge
  University, Wilberforce Road, Cambridge CB3 OWA, UK}
\ead{C.M.Warnick@damtp.cam.ac.uk}

\begin{abstract}
The semi-classical stability of several AdS NUT instantons is
studied. Throughout, the notion of stability is that of stability at
the one-loop level of Euclidean Quantum Gravity. Instabilities
manifest themselves as negative eigenmodes of a modified Lichnerowicz
Laplacian acting on the transverse traceless perturbations. An
instability is found for one branch of the AdS-Taub-Bolt family of
metrics and it is argued that the other branch is stable. It is also
argued that the AdS-Taub-NUT family of metrics are stable. A component
of the continuous spectrum of the modified Lichnerowicz operator on
all three families of metrics is found.
\end{abstract}

\section{Introduction}
Motivated by the AdS/CFT conjecture \cite{Maldacena:1997re}, there has
recently been much interest in the two parameter AdS-Taub-NUT family\footnote{Following \cite{Akbar:2003ed}, we refer to the full two
  parameter family as AdS-Taub-NUT, reserving AdS-Taub-Nut for the
  regular solution containing a nut} of Riemannian
biaxial Bianchi-IX metrics satisfying the Einstein equations with
negative cosmological constant and non-trivial NUT charge
\cite{Akbar:2003ed, Akbar:2002gv, Hartnoll:2005yc, Clarkson:2002uj,Chamblin:1998pz, Hawking:1998ct}. 

Upon the imposition of a regularity condition such metrics divide into two
one parameter classes. The first class (AdS-Taub-Nut) have self-dual Weyl
tensor and contain a nut. The second class of solutions
(AdS-Taub-Bolt) contain a
bolt, this class splits further into two
branches. This is analogous to the case of AdS-Schwarzschild solutions
at a given temperature, where the r\^{o}le of AdS is played by the AdS-Taub-Nut. The AdS-Schwarzshild
solution with a smaller mass (and smaller horizon) is unstable and
it's action is greater than that of both AdS and the other
AdS-Schwarzshild \cite{Hawking:1982dh, Prestidge:1999uq}. We find a similar situation for AdS-Taub-Bolt. 

In \cite{Hartnoll:2005yc}, Hartnoll and Kumar conjectured, based on the
Klebanov-Polyakov version of the AdS$_4$/CFT$_3$ correspondence, that
the global minimiser of the action for the AdS-Taub-NUT class of metrics should be
stable. In \cite{Kleban:2004bv}, the stability of some of these spaces against
scalar perturbations and brane nucleation was discussed. By considering a one-loop correction to the bulk
gravitational partition function, we shall investigate the semi-classical linear
stability of such spacetimes, using techniques developed by Hu
\cite{Hu:1974} and applied to the case of Euclidean Taub-Bolt by Young
\cite{Young:1984dn}. These techniques have also been applied to the
case of Lorentzian Taub-NUT recently by Holzegel \cite{holzegel}. The criterion for instability is that the
modified Lichnerowicz operator
\begin{equation}
\DL h_{ab} = -\nabla^e\nabla_e h_{ab} - 2R_{acbd}h^{cd}
\end{equation}
acting on transverse, trace-free symmetric tensors should have no
negative eigenmodes. We shall investigate the spectrum of this
operator on metric perturbations of the AdS-Taub- instantons which
preserve the $SU(2)$ symmetry. We find a negative mode for one of the
instantons indicating instability for all values of the cosmological
constant. We also find part of the positive spectrum of the modified
Lichnerowicz operator for all instantons in the class under consideration.

In Section 2, we give a brief overview of the
AdS-Taub-Nut and AdS-Taub-Bolt spaces, stating some results which we
will later require. In Section 3, we introduce the method used to
examine the stability of the spaces and we present an instability for
one branch of the AdS-Taub-Bolt spaces and argue that the other spaces
are linearly stable. In section 4 we present an alternative viewpoint,
confirming the claims of section 3. In Section 5 we investigate the
continuous spectrum of the operator $\DL$.

\section{NUTs and Bolts in AdS space}
The metric for both AdS-Taub-Nut and AdS-Taub-Bolt in four dimensions
can be put into the form:
\begin{equation}
\label{metric}
d s^2 = \frac{1}{A(r)} d r^2 + 4 N^2 A(r) \sigma_3^2 + B(r)
(\sigma_1^2 + \sigma_2^2),
\end{equation}
where $\sigma_i$ are the left-invariant $SU(2)$ one-forms. The
functions $A(r)$ and $B(r)$ are given by
\begin{eqnarray}
A(r) &=& \frac{r^2 + N^2 - 2 m r + \ell^{-2} (r^4 - 6 N^2 r^2 - 3
  N^4)}{r^2 - N^2}, \nonumber \\
B(r) &=& r^2 - N^2,
\end{eqnarray}
where $N$ is the NUT charge and $m$ is the mass. This metric is
Einstein, with $R_{\mu \nu} = -\frac{3}{\ell^2} g_{\mu \nu}$. The
fixed point set of the $U(1)$ action is given by the solutions, $r = r_+$ of
$A(r)=0$. If $B(r_+) = 0$, the
fixed point set is of zero dimension and is known as a nut. If $B(r_+)
\neq 0$, the fixed point set is two dimensional and is known as a
bolt. As $r$ approaches $r_+$, we will in general have a conical
singularity unless $A(r)$ satisfies the regularity condition
\begin{equation}
\label{regcond}
\abs{A'(r_+)} = \frac{1}{2 N}.
\end{equation}
This amounts to a relation between $m$ and $N$, which can be solved
and we find that the number of solutions depends on the value of
$\frac{\ell}{N}$. 

For all values of $N$ and $\ell$ there is a solution, called the
AdS-Taub-Nut solution where
\begin{equation}
\label{nutparams}
r_{n} = N, \qquad m_n = \frac{N ( \ell^2 -4 N^2)}{\ell^2}.
\end{equation}
Clearly we have $B(N) = 0$, so we have a regular nut at $r=r_n$. We
can also find two bolt solutions, which I shall call the
AdS-Taub-Bolt$^\pm$ solutions. They are given by:
\begin{eqnarray}
r_{b\pm} = \frac{\ell^2 \pm \sqrt{\ell^4 - 48 N^2 \ell^2 + 144
    N^4}}{12 N},\nonumber \\ m_{b\pm} = \frac{r_{b\pm}^4 + (\ell^2 -
    6 N^2) r_{b\pm}^2 + N^2 (\ell^2 - 3 N^2)}{2 \ell^2 r_{b\pm}}.
\label{boltparams}
\end{eqnarray}
Clearly these equations only make sense if the quantity under the
square root in the first equation is positive. This restricts $\ell/N$ to
the range
\begin{equation}
\label{paramrange}
\frac{\ell}{N} \geq 2 \sqrt{3(2+\sqrt{3})} \approx 6.69.
\end{equation}
Thus we find that requiring that the metric (\ref{metric}) be regular
restricts the freedom quite considerably. For any given value of
$\frac{\ell}{N}$, there are either $1$ or $3$ regular metrics in this
family, excluding the critical case where $r_{\mathrm{b}+} = r_{\mathrm{b}-}$. It is
possible to calculate the action for these spaces, but we shall
postpone this until section 4.

It is useful to consider the limit $\ell \to \infty$, as we expect to
recover the Euclidean Taub-NUT and Taub-Bolt solutions. For the
AdS-Taub-Bolt$^+$ case, this limit does not exist as $r_{\mathrm{b}+} \to \infty$
as $\ell$ gets large. However, for the other two cases we can take
this limit and we find for the nut case:
\begin{equation}
m_n \to N, \hspace{2cm} A(r) \to \frac{r-N}{r+N},
\label{eucnut}
\end{equation}
which gives the well known form of the metric for the self-dual Taub-NUT
instanton. For the AdS-Taub-Bolt$^-$ case, we find:
\begin{equation}
r_{\mathrm{b}-} \to 2 N, \hspace{1cm} m_{b-} \to \frac{5}{4}N, \hspace{1cm} A(r)
\to \frac{r^2 - \frac{5}{2} N r + N^2}{r^2 - N^2},
\end{equation}
which gives the metric for Euclidean Taub-Bolt. This will provide a
useful check on our stability results as for vanishing $\Lambda$,
Taub-NUT has a self-dual Riemann tensor, and hence is linearly stable
\cite{Hawking:1979zs}. In \cite{Young:1984dn} an unstable mode for the
Taub-Bolt instanton was found.

\section{Stability}

\subsection{A criterion for instability}
Instabilities of a physical system may make themselves known as an
imaginary part of the partition function for the system
\cite{Langer:1969bc, Coleman:1978ae}. The partition function for
Euclidean quantum gravity is given by:
\begin{equation}
\label{part}
Z = \int \rmd [g] \rme^{-S[g]},
\end{equation}
where the integral is taken over all Riemannian metrics subject to
appropriate boundary behaviour and periodic in Euclidean time. The
Euclidean action is given by:
\begin{equation}
\label{eucact}
S[g]  = -\frac{1}{16\pi G} \int_{\mathcal{M}} \rmd V
\sqrt{g} \left( R - 2 \Lambda \right) - \frac{1}{8 \pi G}
\int_{\partial \mathcal{M}} \rmd S \sqrt{h}K.
\end{equation}
Unfortunately this is not positive definite. Under a conformal
transformation of the metric, $g^\prime = \Omega^2 g$, where for
simplicity we assume $\Omega = 1$ on a neigbourhood of $\partial \mathcal{M}$ the scalar
curvature transforms like:
\begin{equation}
R^\prime = \Omega^{-2}R - 6 \Omega^{-3} \nabla^a \nabla_a \Omega.
\end{equation}
Thus we find
\begin{equation}
I[g'] = -\frac{1}{16\pi G} \int_\mathcal{M} \rmd V \sqrt{g}  \left ( \Omega^2 R +6 \Omega_{,a}
\Omega^{,a} \right ) - \frac{1}{8\pi G} \int_{\partial
  \mathcal{M}} \rmd S\sqrt{h} K.
\end{equation}
So by taking $\Omega$ to vary quickly, we can make $I[g']$ as negative
as we choose. This problem can be circumvented by an appropriate
choice of contour, but only in the
semi-classical (one-loop) approximation \cite{Gibbons:1978ac}.
For a semi-classical approximation, we expand the integral
(\ref{part}) about the critical points of $S[g]$, where we expect the
dominant contributions to the integral. These satisfy
\[
\frac{\delta S}{\delta g_{a b}} = 0 \rightarrow R_{ab} = \Lambda g_{ab},
\]
i.e., the classical Riemannian vacuum Einstein equations. Thus in order to
find the semi-classical approximation to $Z$, one expands in small perturbations about the
classical solutions:
\begin{eqnarray}
\tilde{g}_{ab} &=& g_{ab} + h_{ab}, \nonumber \\
S[\tilde{g}] &=& S_0[g] + S_2[h] + O(h^3),
\end{eqnarray}
where $S_2[h]$ is second order in the small perturbation $h_{ab}$. We truncate the series for $I[\tilde{g}]$ and integrate over $h_{ab}$
to get
\begin{equation}
\label{fint}
Z \cong Z_{1\hspace{0.05cm} \mathrm{loop}} = N\rme^{-S_0[g]} \int \rmd [h] \rme^{-S_2[h]},
\end{equation}
where the integral is to be taken over physically distinct
perturbations $h_{ab}$. The action $S_2[h]$ is invariant under gauge transformations which
correspond to infinitesimal diffeomorphisms:
\[
h_{ab}' = h_{ab} + \del{a}V_b + \del{b}V_a.
\]
In order to deal with this, we follow \cite{Gibbons:1978ji} and
use the Fadeev-Popov gauge fixing technique. The gauge fixing
condition will be:
\begin{equation}
\label{gauge}
\del{a} \left ( h^{ab} - \frac{1}{\beta} g^{ab} h \right) = W^b,
\end{equation}
where $W$ is an arbitrary vector, $\beta$ is an arbitrary constant and
$h = h^a{}_a$ is the trace of the perturbation. The standard
Fadeev-Popov method allows us to re-write the functional integral
(\ref{fint}) as an integral over all fields $h_{ab}$, but with an
altered integrand:
\begin{equation}
\label{1lp}
 Z_{1\hspace{0.05cm} \mathrm{loop}}= N \rme^{-S_0[g]} \int \rmd[h] (
 \mathrm{det}\ C) \rme^{-I_{2 \hspace{0.05cm} \mathrm{eff.}}[h]}.
\end{equation}
We now decompose the metric perturbation into a transverse, traceless
part $\phi_{ab}$, a longitudinal traceless part generated by the vector $\tilde{n}$ and
the trace $h$. Further, a Hodge-de Rham decomposition can be performed
on $\tilde{n}$ so that
\[
\tilde{n}_a = \eta_a + \del{a} \chi, \hspace{0.5cm} \hbox{with} \hspace{0.5cm} \del{a} \eta^a =  0.
\]
We can then write the effective action as:
\begin{equation}
I_{2 \hspace{0.05cm} \mathrm{eff.}} = -\frac{1}{16 \pi} \int \rmd V \mathcal{L}_2,
\end{equation}
with
\begin{equation}
\fl \mathcal{L}_2 = -\frac{1}{4}\phi^{ab}G_{abcd}\phi^{cd} +
\frac{1}{2}hFh - 2 \gamma D^{ab}(\chi)D_{ab}(F\chi) - \gamma
\mathfrak{D}^{abe}(\eta_e) \mathfrak{D}_{abf}((C_2)^{fd}\eta_d)
\end{equation}
where $\gamma$ is a constant related to $\beta$ by $\beta = 4 \gamma /
(1+\gamma)$ and we have defined operators
\begin{eqnarray}
F &=& \frac{1-3\gamma}{16\gamma} \del{e}\nabla^e - \frac{1}{4}\Lambda,
\nonumber \\
D_{ab} &=& 2 \del{a} \del{b} - \frac{1}{2}g_{ab}\del{e} \nabla^e,
 \nonumber \\
\mathfrak{D}_{abe}(V^e) &=& \frac{1}{2} \left(\del{a}V_b +\del{b}V_a
\right), \nonumber \\
(C_2)_{ab}V^b &=& -\left(\del{e} \nabla^e + \Lambda \right)V_a,\nonumber \\
G_{abcd} \phi^{cd} &=& -\del{e}\nabla^e \phi_{ab} - 2R_{acbd}\phi^{cd},
\end{eqnarray}
where $V$ is a divergence free vector. It is possible to write the
$(\mathrm{det}\ C)$ factor in (\ref{1lp}) as a functional integral over
anti-commuting vectors. Applying a Hodge-de Rham decomposition one
finds
\[
(\mathrm{det}\ C) \sim (\mathrm{det}\ C_2)(\mathrm{det}\ F).
\]
Finally, evaluating the Gaussian integrals in (\ref{1lp}) we find
\begin{equation}
Z_{1\hspace{0.05cm} \mathrm{loop}} \sim (\mathrm{det} \
G)^{-\frac{1}{2}} (\mathrm{det}\ C_2)^{\frac{1}{2}}.
\label{dets}
\end{equation}
Any zero modes of the determinants in (\ref{dets}) should be projected
out. The determinants can be regularised by a $\zeta$-function
regularisation \cite{Gibbons:1978ji}. $C_2$ is a positive
semi-definite operator on the space of divergence free vector fields,
with zero modes corresponding to Killing vectors. $G$ however is not
positive definite and can in fact have a negative eigenmode. Such a
negative eigenmode would introduce an imaginary part to the partition
function and thus herald an instability. For example the flat
Schwarzschild solution has a negative eigenmode. It is to be expected
that the partition function be pathological in this case since the canonical
ensemble for black holes breaks down due to the fact that they have
negative specific heat.

Thus in our search for instabilities we can restrict to perturbations
$h_{ab}$ which are transverse and tracefree. Our criterion for
instabilities is that there exist negative eigenvalue solutions to the
eigenvalue equation:
\begin{equation}
\label{licn}
(\Delta_L h)_{ab} - 2 \Lambda h_{ab} = -\nabla_e \nabla^e h_{ab} -
2R_{acbd} h^{cd} = \lambda h_{ab},
\end{equation}
where we have re-expressed $G$ in terms of the Lichnerowicz Laplacian
$\Delta_L$ \cite{Lichnerowicz:1961}. This equation is consistent with
the Transverse Traceless condition.

A negative eigenmode of (\ref{licn}) corresponds to a direction within
the space of gauge fixed metric perturbations along which the
classical solution $g_{ab}$ is a local maximum. The criterion that
\[
\widetilde{\Delta}_L~=~\Delta_L~-~2\Lambda~\geq 0
\]
for stability may be arrived at from a more geometric perspective (see
for example Besse \cite{Besse}) which avoids the Fadeev-Popov
procedure used above. There are some cases in which it can be shown
that $\widetilde{\Delta}_L$ is positive definite. These include the
case when $g_{ab}$ has self-dual Riemann tensor (sometimes called the
half-flat condition) \cite{Hawking:1979zs} and also the case where
$g_{ab}$ is Einstein-K\"{a}hler \cite{Pope:1982ad}. In both these
cases, the existence of a covariantly constant spinor is used to
relate the eigenmodes of $\widetilde{\Delta}_L$ to those of scalar
Laplacians acting on charged fields.

\subsection{Hu's Technique}

Hu \cite{Hu:1974} proposed a method for
separation of variables of tensor equations in a homogeneous space
where the metric can be written in the form
\begin{equation}
\label{hu}
d s^2 = d r^2 + \overline{\gamma_{ij}} \sigma^i \sigma^j,
\end{equation}
where $i, j = 1,2,3$ and $\sigma^i=\sigma_i$ are again the left
invariant one-forms on $SU(2)$. This was generalised by Young
\cite{Young:1984dn} to the case where $g_{00}$ is permitted to depend
on $r$. The metric perturbation is expanded in terms of the Wigner
functions, $\mathcal{D}_{KM}{}^J(\theta, \phi, \psi)$ which are the
analogue on $S^3$ of the spherical harmonics $Y_{l, m}(\theta, \phi)$. We concentrate on
diagonal\footnote{ The equations for off-diagonal perturbations can be put into
Schr\"{o}dinger form with strictly positive potential (see
(\ref{schro})), for all three
metrics considered here and so cannot give rise to negative eigenmodes.}
metric perturbations with the lowest ``angular momentum'',
$J=M=K=0$. These are the perturbations which preserve $SU(2)$
symmetry. The resulting equations, along with the transverse and traceless
conditions for the general metric form (\ref{metric}) are
given in the Appendix.

We first consider the $11$ and $22$ equations. Taking the difference
of (\ref{11cpt}) and (\ref{22cpt}) we find a second order differential
equation for $Y = h_{11} - h_{22}$. We can put this into
Schr\"{o}dinger form:
\begin{equation}
\label{schro}
-\frac{d^2 \chi}{d r_*^2} + V(r_*) \chi = \lambda \chi, \hspace{1.5cm}
 0 \leq r_*< \infty, 
\end{equation}
via the substitutions $Y = f \chi$ and $dr = g dr_*$, for some
suitable $f$ and $g$. When we do this, we find that $V(r(r_*))$ is a
positive function for $r > r_+$, so that no normalizable solutions to
(\ref{schro}) exist with $\lambda <0$. This means that in the search for a negative
eigenmode of (\ref{licn}) we may set $h_{11} = h_{22}$.

We now consider the other diagonal mode. Setting $h_{11} =h_{22}$, we
can use the constraint equations (\ref{tracefree}), (\ref{transverse})
with the 00 equation (\ref{00cpt}) to decouple a second order
differential equation in $h_{00}$ from the others. If we can solve
this, we can find $h_{11}$, $h_{22}$ and $h_{33}$ from the constraint
equations. Since (\ref{licn}) is consistent with the transverse
traceless condition, we know that (\ref{tracefree}),
(\ref{transverse}) and (\ref{00cpt}) imply (\ref{11cpt}) and
(\ref{33cpt}). This can be checked explicitly. The equation we
decouple may be written:
\begin{equation}
\label{diffeq}
a(r) h''_{00} + b(r) h'_{00} + c(r) h_{00} = 0,
\end{equation}
where differentiation with respect to $r$ is denoted by a prime and
the coefficients are given by: 
\begin{eqnarray*}
a(r) &=& A, \\
b(r) &=& 3 A' + \frac{AB'}{B} + \frac{B(A'^2 - 2A
  A'')}{B A' - A B'} - \frac{A}{B} \left( \frac{2AB'^2 -
  BA'B' - 2ABB''}{B A' - A B'} \right), \\
c(r) &=& \lambda + \frac{A'^2}{2A} + \frac{A'B'}{B} -
\frac{AB'^2}{B^2} + A'' + \frac{3}{2A} \left( \frac{BA'
  + A B' }{BA' - AB'}\right) \\ && \hspace{.5cm} + \frac{B A'B' +
  2BAB'' - 2AB'^2}{B^2} \left( \frac{2BA'
  + A B' }{BA' - AB'}\right),
\end{eqnarray*}
the rest of this section will consist of an analysis of this equation.

\subsection{AdS-Taub-Nut}

In the case where $A(r)$ and $B(r)$ are given by (\ref{metric}) with
(\ref{nutparams}) we can cast equation (\ref{diffeq}) into
Schr\"odinger form (\ref{schro}) where $V$ is once again found to be
positive on the range $r>r_+$. Thus for AdS-Taub-Nut we have checked
all the possible TT perturbations with $J=K=M=0$ and found that none
give negative eigenvalue solutions to (\ref{licn}), thus there is no
linear instability due to perturbations of this type. This is in
agreement with the result that the $\ell \to \infty$ limit yields a
(linearly) stable metric.

One would normally expect the eigenvalues of a Laplacian operator to
be bounded below by the most symmetric modes. In our case that would
be the $SU(2)$ invariant modes. In the case of the scalar Laplacian
acting on a space with metric given by (\ref{metric}) this can be seen
explicitly:
\begin{eqnarray}
-\nabla^2 \phi &=& -\nabla^2\left(\phi^K(r)\mathcal{D}_K(\theta,
 \phi,\psi)\right) \nonumber\\ &=& \left(-\nabla^2 \phi^K(r)\right)\mathcal{D}_K(\theta,
 \phi,\psi) - 2 \left(\partial^d \phi^K(r)\right)\left(\partial_d\mathcal{D}_K(\theta,
 \phi,\psi)\right) \nonumber\\ && \quad - \phi^K(r) \nabla^2\mathcal{D}_K(\theta,
 \phi,\psi) \nonumber \\
&=& \left(-\nabla^2 \phi^K(r)\right)\mathcal{D}_K(\theta,
 \phi,\psi) + \phi^K(r)\left(- \nabla^2\mathcal{D}_K(\theta,
 \phi,\psi)\right). \label{sclap}
\end{eqnarray}
It can be shown that $(-\nabla^2\mathcal{D}_K)$ is positive for this
metric form. This separation into a radial part and an angular part
does not occur for the the Laplacian acting on symmetric tensors as
there are extra terms introduced by the connection. It has not been
possible to show explicitly that the $SU(2)$ invariant perturbations
give a lower bound for $\DL$, but one would still expect this to be
the case. Increasing the spin of the perturbation modes introduces a
more rapidly varying angular dependence, which one would expect to
increase the eigenvalues of $-\nabla^2$ and hence of $\DL$. 

We therefore conjecture that $\DL$ is positive acting on all metric
perturbations and thus that AdS-Taub-Nut is stable. This may be
related to the fact that the Weyl tensor is self dual for the
AdS-Taub-Nut space. In the Euclidean case, the full Riemann tensor is
self-dual and this is known to imply stability \cite{Hawking:1979zs}.

\subsection{AdS-Taub-Bolt$^-$}

In the case of AdS-Taub-Bolt$^-$, when one changes (\ref{diffeq}) into
Schr\"{o}dinger form, one encounters a divergent potential. In order to
proceed, we must use numerical integration techniques. If we make the
substitution $r = x N$, we find that the equations depend on $N$
only through the ratio $\frac{\ell}{N}$. We can therefore scale $N$
out of the problem and set $N=1$ without loss of generality. We must
now consider the singularity structure of the differential equation
(\ref{diffeq}). In the region of interest to us, $r \geq r_{\mathrm{b}-}$, we
find that the equation has 3 regular singular points, at $r_{\mathrm{b}-}$,
$r_{\mathrm{int.}}$ and infinity, where $r_{\mathrm{b}-} < r_{\mathrm{int.}} < \infty$. Other
singular points outside of the range of interest preclude the possibility
of an analytic solution.

We consider first the regular singular point
at $r = r_{\mathrm{b}-}$. Near to this point, the differential equation has the
asymptotic form: 
\begin{equation}
\label{rbpeq}
h''_{00} + \frac{4}{r-r_{\mathrm{b}-}}h'_{00} + \frac{2}{(r-r_{\mathrm{b}-})^2}
h_{00} = 0.
\end{equation}
Substituting $h_{00} \sim (r-r_{\mathrm{b}-})^\alpha$, we find the indicial equation
\[
\alpha^2 + 3 \alpha + 2 = 0,
\]
with solutions $\alpha = -1$ and $\alpha = -2$. Thus there are two
independent solutions of (\ref{diffeq}) that behave like $(r-r_{\mathrm{b}-})^{-1}$
and $(r-r_{\mathrm{b}-})^{-2}$ as $r \to r_{\mathrm{b}-}$. We impose a
normalisation condition on the perturbation modes given by
\begin{equation}
\label{norm}
\int \rmd V h_{ab} h^{ab} < \infty.
\end{equation}
Using the constraint equations, we can show that the $h_{00}$
contribution dominates the integrand in a
region just outside the bolt. Using the fact that $dV = \sqrt{g}d
\psi d\theta d \phi d r$, where $\psi$, $\theta$ and $\phi$ are the
usual Euler angles on $SU(2)$ and the form of the metric (\ref{metric}) with
the constraint equations (\ref{tracefree}, \ref{transverse}) we
can show that this integral coverges at $r=r_{\mathrm{b}-}$ if and only if
$h_{00} \sim (r-r_{\mathrm{b}-})^\beta$ where $\beta > -3/2$. Thus only one of
the two independent solutions at $r = r_{\mathrm{b}-}$ represents a
normalisable perturbation -- we must pick the solution that behaves like
$h_{00} \sim (r-r_{\mathrm{b}-})^{-1}$ near $r = r_{\mathrm{b}-}$. This provides us with a
stepping off condition for our numerical integration. We can perform a
Frobenius expansion of the solution about the regular singular point
and then use this to calculate $h_{00}$ and $h'_{00}$ at $r =
r_{\mathrm{b}-} + \epsilon$ where $\epsilon$ is some small number, taken to be
$10^{-5}$. This deals with the first regular singular point.

It can be shown, by a similar method to that used above that $h_{00}$ and
$h'_{00}$ are bounded as $r \to r_{\mathrm{int.}}$, so no special
considerations are required for the numerical integration through this point.

Finally we need to take account of the regular singular point at
infinity. In the limit that $r \to \infty$, the differential equation
has the asymptotic form
\begin{equation}
\label{infeq}
h''_{00} + \frac{12}{r} h'_{00} + \frac{28 + \ell^2
  \lambda}{r^2} h_{00} =0.
\end{equation}
Substituting $h_{00} \sim r^\alpha$ we find the indicial equation:
\[
\alpha^2 + 11 \alpha + (28+\ell^2 \lambda) = 0,
\]
with solutions
\begin{equation}
\label{alp}
\alpha_{\pm} = \frac{-11 \pm \sqrt{9-4\ell^2 \lambda}}{2}.
\end{equation}
Using the constraint equations, we find that the integrand of
(\ref{norm}) is dominated by the contribution from $h_{33} \sim r^6
h_{00}$. This translates to a requirement that $h_{00} \sim r^\beta$
with $\beta < -\frac{11}{2}$ in order that (\ref{norm}) is
satisfied. So we see that only one of the two independent solutions at
infinity will give a normalisable perturbation. We will need to match
the normalisable solution at infinity to the normalisable solution
at $r_{\mathrm{b}+}$ and this matching, if it is possible, will determine the
value of $\lambda$.

In order to ascertain whether there exist negative solutions for
$\lambda$, we perform a numerical integration starting at
$r_{\mathrm{b}-}+\epsilon$, with initial conditions determined from the
Frobenius expansion about $r_{\mathrm{b}-}$. This determines the
solution, up to a arbitrary multiplicative constant as we can determine $h_{00}$ and all its derivatives
here. Having chosen the normalisable solution at
the bolt, we need to check that the solution is normalisable at
infinity. The only parameter we still have available is $\lambda$. As
$r \to \infty$ the solution must have the asymptotic form found above,
but will in general be a linear combination of the two independent asymptotic
forms.
\begin{equation}
\label{asymp}
h_{00}(r) \sim k(\lambda) r^{\alpha_+} + l(\lambda) r^{\alpha_-},
\end{equation}
where $\alpha_+>-11/2$ and $\alpha_- < -11/2$ are given in equation
(\ref{alp}) and $k(\lambda), l(\lambda)$ are constants, which we
assume to depend continuously on $\lambda$. The normalisation condition requires that $\lambda =
\lambda_0$, where $k(\lambda_0) =
0$. Since we are only performing a numerical integration, it is not
possible to find the form of $k(\lambda)$ explicitly, but we can find
an interval within which $\lambda_0$ must lie. 

If we consider $f(r) = r ^{11/2}
h_{00}(r)$, then $f(r)$ will generically tend to either positive or
negative infinity, depending on the sign of $k(\lambda)$. If we can
find numbers $\lambda_1$ and $\lambda_2$ such that $k(\lambda_1)$ has
opposite sign to $k(\lambda_2)$, then we can deduce the existence of a
root of $k(\lambda)$ in the interval $(\lambda_1, \lambda_2)$, by the
Intermediate Value Theorem. This procedure was implemented using the
computer package Mathematica, and negative
eigenvalues have been found for values of $\ell$ on the entire range
$2 \sqrt{3(2+\sqrt{3})} < \ell < \infty$. A plot of these
negative eigenvalues is shown in Figure 1, together with a best fit curve.

\begin{figure}[!t]
\centering \framebox {\includegraphics[height=3in,
width=5in]{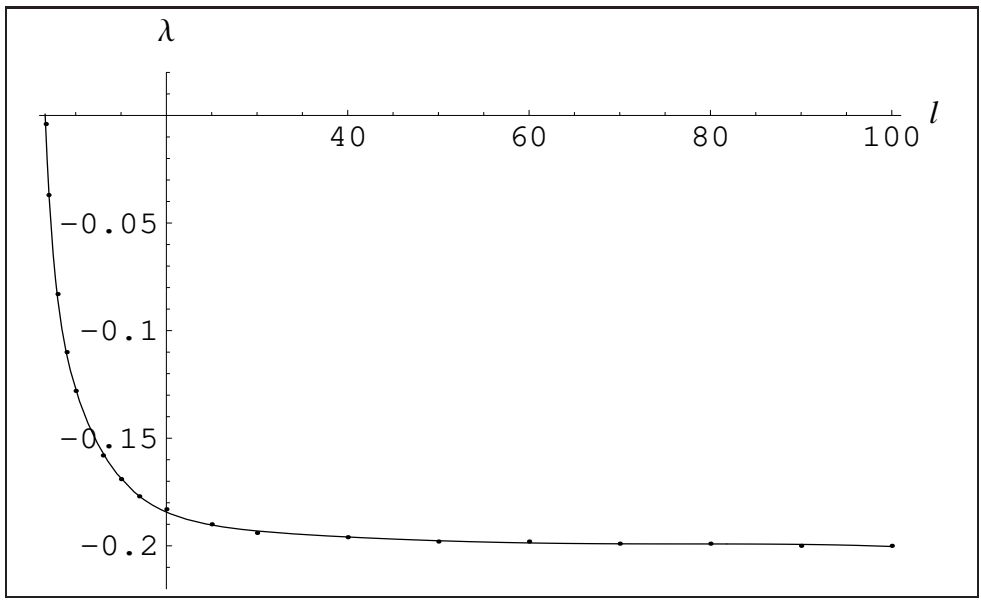}}
\caption{A plot showing the negative eigenvalue, $\lambda$, against
  $\ell$ for AdS-Taub-Bolt$^-$ Space}
\end{figure}

We find that as $\ell \to \infty$, we have that $\lambda \to -0.20$
approximately. This is consistent with the findings of Roberta Young \cite{Young:1984dn},
who found that Euclidean Taub-Bolt had a negative eigenmode in this
sector with $\lambda \approx -0.20$. We also find that as $\ell$
approaches the critical value where the AdS-Taub-Bolt$^-$ and
AdS-Taub-Bolt$^+$ spaces are the same, the negative eigenvalue tends to
zero from below.

\subsection{AdS-Taub-Bolt$^+$}

The final case, where the bolt is located at $r = r_{\mathrm{b}+}$ is similar
in many ways to the previous case. We find the same singularity
structure and asymptotic forms for the perturbation. We can proceed in
exactly the same way as above, but we find no negative eigenvalues. We
therefore conjecture that this spacetime is stable, as it appears to
have no negative eigenmodes in the lowest ``angular momentum'' state.

\begin{figure}[!h]
\centering \framebox {\includegraphics[height=3in,
width=5in]{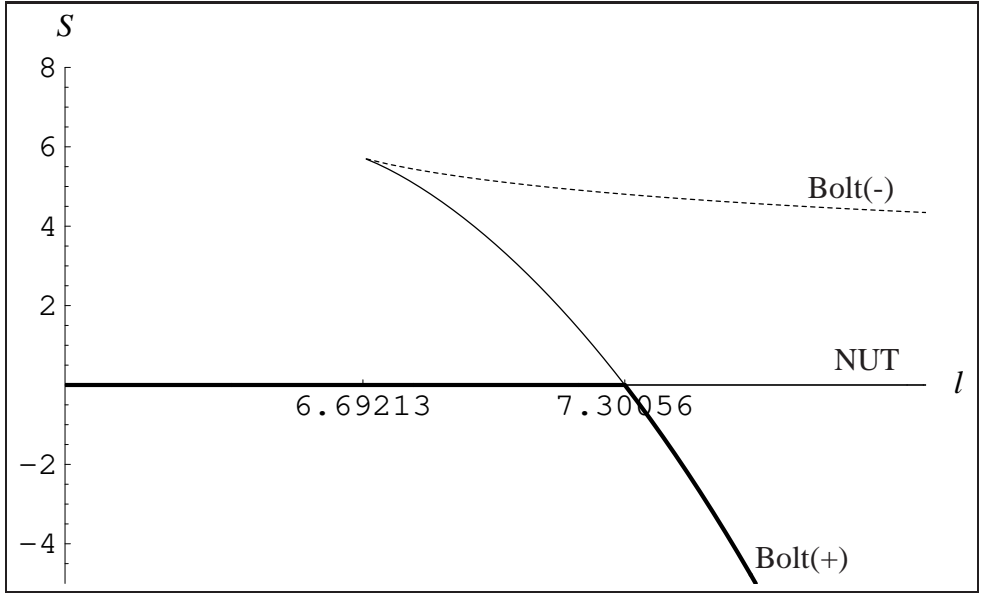}}
\caption{A plot showing the actions $S-S_{NUT}$ against $\ell$ for
  $N=1$}
\begin{picture}(1,1)(0,0)
\put(120,185){Bolt(-)}
\put(130,120){NUT}
\put(100,45){Bolt(+)}
\end{picture}
\end{figure}

In Figure 2 we show a plot of the action of both AdS-Taub-Bolt spaces
against $\ell$, with the action for AdS-Taub-Nut subtracted. This will be calculated in Section 4. The dotted line
represents the unstable AdS-Taub-Bolt$^-$ space, the horizontal axis
the Taub-NUT space and the other curve AdS-Taub-Bolt$^+$. The thicker
line denotes the global minimizer of the action. We see here
two interesting features. Firstly we note that at $\ell = 2
\sqrt{3(2+\sqrt{3})}$ we have a bifurcation, with the two Bolt
families appearing, with one stable and the other unstable. We also
note that at $\ell = 2 \sqrt{7 + 2\sqrt{10}}$ there is a phase
transition from AdS-Taub-Nut as the global minimiser of the
action to AdS-Taub-Bolt$^+$ as the global minimiser. This is the NUT
charged version of the Hawking-Page transition for AdS Black Holes \cite{Hawking:1982dh}.

\section{Another Perspective on Stability}

We will now present an alternative argument, which reproduces all the
features of the plot in Figure 2.

In constructing the nut charged AdS instantons considered above, the usual
technique (outlined in Section 2) is to consider a metric of the form
(\ref{metric}) which satisfies the Einstein equations $R_{\mu \nu} =
-\frac{3}{\ell^2}g_{\mu \nu}$. Then one imposes the condition of
regularity to reduce the number of solutions to those given
above. Here we shall consider a family of metrics which are regular
everywhere, with a bolt (or possibly nut) located at $r_b$, which is
treated as a free parameter. The AdS-Taub-Nut and Taub-Bolt($\pm$)-AdS
instantons correspond to particular choices of $r_b$. We can then
calculate the action for this family of spacetimes as a function of
$r_b$ and we expect it will be locally extremized at the already known
values of $r_b$ which give solutions of the Einstein equations. This
corresponds to taking a slice parameterised by $r_b$ through the space
of metrics containing a bolt or a nut.

Our metric ansatz will be:
\begin{equation}
\label{metric2}
d s^2 = \frac{1}{A(r)} d r^2 + 4 N^2 A(r) \sigma_3^2 + B(r)
(\sigma_1^2 + \sigma_2^2),
\end{equation}
with
\begin{eqnarray*}
A(r) &=& \frac{(r - r_b)(\alpha r + \beta + \ell^{-2}(r^3 + r^2 r_b))}{r^2-N^2}, \\
B(r) &=& r^2 - N^2.
\end{eqnarray*}
We then impose regularity at $r = r_b$ and require the mass to be $m$,
by considering the asymptotics. This
gives us the metric functions:
\begin{eqnarray}
\fl A(r) &=& \frac{\left( r - r_b \right) \left( 4N\left( r - r_b \right) r_b
       {\left( r + r_b \right) }^2 - \ell^2 \left( 4mN \left( r_b -r \right)  + 
         \left( r + rb \right) \left( n^2 - r_b^2 \right)  \right)
       \right) }{4 \ell^2N\left( r^2 - N^2 \right)
   r_b} , \nonumber\\
\fl B(r) &=& r^2 - N^2. \label{ans}
\end{eqnarray}
We are now ready to calculate the action. The Euclidean action is
given by:
\begin{equation}
\fl S = S_{\mathrm{bulk}} + S_{\mathrm{surf.}} = -\frac{1}{16\pi G}
\int_{\mathcal{M}} \rmd V
\sqrt{g} \left( R + \frac{6}{\ell^2} \right) - \frac{1}{8 \pi G}
\int_{\partial \mathcal{M}} \rmd S \sqrt{h}K,
\label{actionint}
\end{equation}
which is the standard Einstein-Hilbert action together with
the Gibbons-Hawking-York boundary term. It is well known that this
integral does not converge and so some means of regularising the
action is required. The method used was that of counterterm
subtraction proposed by Emparan, Johnson and Myers
\cite{Emparan:1999pm}. This was also found independently by Mann \cite{Mann:1999pc}. The calculation proceeds much as their
calculation for the action of Taub-NUT. We find that the action of the
metric defined by (\ref{metric2}) and (\ref{ans}) is:
\begin{equation}
\label{action}
 S = \frac{4N}{\ell^2}\left( r_b^3 - 3N^2 r_b \right ) + N^2 + 4Nr_b - r_b^2.
\end{equation}
We note that this does not depend on $m$, which is also a free
parameter of this family of metrics. The variation of this
reduced action alone will not determine $m$, for that we require the
full Einstein equations, which force $m$ to take the NUT or Bolt
values at the critical points. Setting $r_b = x N$,defining
$\mu = \ell/N$, and using
the additive freedom to set $S=0$ for AdS-Taub-Nut we find
\[
\frac{S}{N^2} = \frac{4}{\mu^2} (x^3-3x+2) -3 +4x - x^2.
\]
\begin{figure}[!ht]
\centering \framebox {\includegraphics[height=3in,
width=5in]{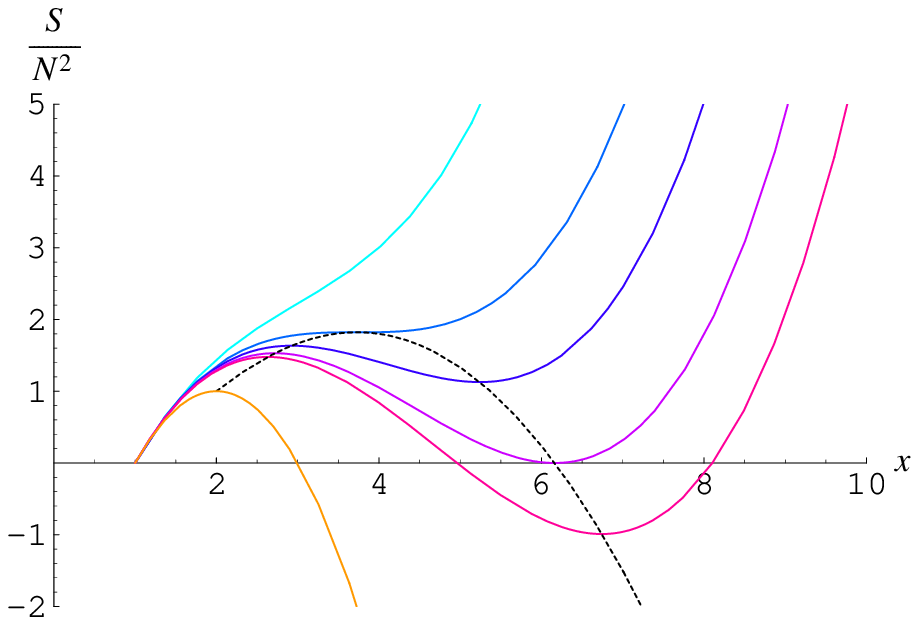}}
\caption{A plot showing $S/N^2$ against $x$ for several values of
  $\mu=\ell/N$}
\begin{picture}(1,1)(0,0)
\put(0,225){\scriptsize $\mu=6$}
\put(45,225){\scriptsize $\mu=\mu_1$}
\put(80,225){\scriptsize $\mu=7$}
\put(110,225){\scriptsize $\mu=\mu_2$}
\put(140,200){\scriptsize $\mu=7.5$}
\put(-45,40){\scriptsize $\mu=\infty$}
\end{picture}
\end{figure}

So we see that as $\mu$ is varied, the graph of $S/N^2$ will
change. A plot of $S/N^2$ against $x$ for various values of $\mu$
is shown in Figure 3. Also included is a dotted curve showing the
locus of the extremal points of the function as $\mu$ varies. We see
that the point $x=1$, which corresponds to AdS-Taub-Nut is always a
minimum since we must exclude the region $x<1$ from consideration as
these metrics are not regular. For small $\mu$ this is the only
minimum. As we increase $\mu$, we find that a pair of extrema, one
maximum and one minimum appear at $\mu = \mu_1$, where
\begin{equation}  
\label{mu1}
\mu_1 =   2 \sqrt{3(2+\sqrt{3})},
\end{equation}
which is precisely the value we would expect from the standard
constructions for AdS-Taub-Nut and AdS-Taub-Bolt. If we think of this
function $S(x)/N^2$ as defining the dynamics of some system by
\[
\ddot{x} = -S'(x)/N^2,
\]
then this bifurcation is of the form known as a saddle-node
bifurcation, since it produces two new fixed points, one stable (a
node) and one unstable (a saddle). Of course we have no formal
dynamics on the space of metrics, but we still expect a local maximum
to correspond to an unstable metric as it corresponds to a direction
contributing an imaginary part to the partition function. Thus we
shall refer to this bifurcation as a saddle-node
bifurcation. We also find that the maximum occurs at precisely
$x=r_{\mathrm{b}-} /N$ and the minimum at $x=r_{\mathrm{b}+}/N$, confirming our previous
result that we expect the AdS-Taub-Bolt$^-$ instanton to be unstable
while the AdS-Taub-Bolt$^+$ instanton is stable.

We also note a global bifurcation  when $\mu = \mu_2$ where
\begin{equation}
\label{mu2}
\mu_2 =  2 \sqrt{7 + 2\sqrt{10}},
\end{equation}
when the global minimum moves from $x=1$ to $x=r_{\mathrm{b}+}/N$, which
corresponds to AdS-Taub-Bolt$^+$. This again accords with what we
expect from previous analysis and corresponds to the Hawking-Page type
phase transition. 

These observations accord nicely with the general
argument given in section 6 of \cite{Gibbons:2002th} for the
production of negative modes of the Lichnerowicz operator at
bifurcation points in parameter space, in particular, from figure 1 we
find that the negative eigenmode tends to zero as we approach the bifurcation.

Unfortunately this analysis cannot replace that of the previous
section, since we have not explicitly found a normalisable negative
mode of (\ref{licn}) for a perturbation about the AdS-Taub-Bolt$^-$
instanton. It is possible to find the linearised perturbation of the
metric at this point using our analysis, but this perturbation is not normalisable. It is
however not in the Transverse, Tracefree gauge and it is possible that
in the appropriate gauge this perturbation does give a
normalisable mode.

\section{The Continuous Spectrum of $\DL$}

We have been so far concerned with looking for eigenvalues of $\DL$ on
the AdS-Taub-Nut and -Bolt instantons. In the case of an elliptic
operator on a non-compact manifold, the spectrum may also include a
continuum of approximate eigenvalues. As an example, the time independent Schr\"{o}dinger operator
for the Hydrogen atom is a Laplacian type operator acting on
$\mathbb{R}^3/\{0\}$. As is well known, this has a countable set of
negative eigenvalues corresponding to the bound states. The spectrum
however includes a continuum of positive ``eigenvalues'' which
correspond to free particles moving in the Coulomb potential. These
are not true eigenvalues since they do not correspond to normalisable
eigenfunctions (wavefunctions). The sense in which they may be
considered part of the spectrum is given below. We shall show that
the continuous spectrum of $\DL$ acting on any of the three spaces
considered above includes the ray $\{\lambda \in \mathbb{R} : \lambda>
9/4l^2\}$.

\subsection{A Little Functional Analysis}

We can think of the space of gauge fixed, finite metric perturbations
as a Hilbert space, defined by:
\begin{equation}
\fl \mathcal{H}=\{h_{ab} \in T^*\mathcal{M} \otimes_S T^*\mathcal{M} :
\nabla^ah_{ab}=0, h^a{}_a=0, \int_{\mathcal{M}}h_{ab} h^{ab} \rmd V<\infty \},
\end{equation}
with $L^2$ inner product given by
\begin{equation}
\langle h,k\rangle=\int_{\mathcal{M}} h_{ab}k^{ab}\rmd V .
\end{equation}
$\DL$ is then a linear operator from $\mathcal{H}_2$ to
$\mathcal{H}_0$, where by $\mathcal{H}_i$ we mean
\begin{equation}
\mathcal{H}_i=\{h_{ab}\in\mathcal{H}:\partial_{c_1}\ldots\partial_{c_i}h_{ab} \hspace{0.2cm}
\makebox{exist and are continuous} \}.
\end{equation}
$\mathcal{H}_i$ are dense subsets of $\mathcal{H}$ with respect to the
norm topology, where the norm is defined as usual by:
\begin{equation}
\norm{h}^2 = \langle h,h \rangle =  \int_{\mathcal{M}}h_{ab}h^{ab}\rmd
V.
\end{equation}
A linear operator $A : \mathcal{H} \to \mathcal{H}$ is said to be bounded if $\exists$ $k$
such that
\begin{equation}
\norm{Ax} \leq k \norm{x}, \qquad \forall x \in \mathcal{H},
\end{equation}
an operator which is not bounded is unbounded. We define the spectrum
of $A$, $\sigma(A)$ as follows:

\vspace{0.8cm}
$\lambda \notin \sigma(A) \iff$ the following properties hold:
\begin{enumerate}
\item[(i)] $(A-\lambda I)^{-1}$ exists,
\item[(ii)] $(A-\lambda I)^{-1}$ is bounded,
\item[(iii)] $(A-\lambda I)^{-1}$ is densely defined (i.e.\ defined on a
  dense subset of $\mathcal{H}$).
\end{enumerate}
\vspace{0.8cm}
The set of $\lambda$ such that $(A-\lambda I)^{-1}$ does not exist is
called the point spectrum of $A$, $P\sigma(A)$. the set of $\lambda$ such that
$(A-\lambda I)^{-1})$ is unbounded is called the continuous spectrum
of $A$, $C\sigma(A)$. We shall not discuss the third condition.

\begin{lemma}
Given $\lambda \notin P\sigma(A)$, if there exists a sequence $x_n \in
\mathcal{H}$ such that
\[
\frac{\norm{(A-\lambda I)x_n}}{\norm{x_n}} \to 0, \qquad
\mathrm{as} \qquad n \to \infty,
\]
then $\lambda \in C\sigma(A)$.
\label{lem}
\end{lemma}
\begin{proof}
Define $y_n =  (A-\lambda I)x_n$, by assumption as $n \to \infty$
\[
\frac{\norm{y_n}}{\norm{(A-\lambda I)^{-1}y_n}} \to 0
\qquad\Rightarrow\qquad \frac{\norm{(A-\lambda I)^{-1}y_n}}{\norm{y_n}} \to \infty,
\]
but this clearly cannot occur if $(A-\lambda I)^{-1}$ is bounded,
since if this is the case, $\exists$ $k$ such that
\[
 \frac{\norm{(A-\lambda I)^{-1}y_n}}{\norm{y_n}} \leq k.
\]
Thus $(A-\lambda I)^{-1}$ must be unbounded, hence $\lambda \in
C\sigma(\lambda)$.
\end{proof}

\subsection{The $\DL$ Operator}

We now consider the continuous spectrum of $\DL$. We will restrict
attention for the moment to perturbations of the form considered in
section 3. These are generated by the $h_{00}$ component, which has to
satisfy equation (\ref{diffeq}). We first define the functions
$\rho_n(x)$ by:
\begin{equation}
\rho_n(x) = \left\{ \begin{array}{ll}
 1 & \textrm{if $r<n$}\\
 1-K\int_n^r \exp \left [-\frac{1}{(x-n)^2}-\frac{1}{(x-n-1)^2} \right
 ] \rmd x & \textrm{if
 $n\leq r \leq n+1$ }\\
 0 & \textrm{if $r>n+1$}
  \end{array} \right.
\label{rhon}
\end{equation}
with $K$ chosen so that $\rho_n(n+1)=0$. These functions are everywhere smooth on the real line. For $\lambda
>9/4l^2$ we define $f(r)$ to be the solution of (\ref{diffeq}) which
gives a finite contribution to $\langle h ,h \rangle$ at $r_+$. This
generates a perturbation which satisfies
\[
(\DL-\lambda I)h_{ab} = 0.
\]
However this $h_{ab}$ is not an element of $\mathcal{H}$ since as $r \to \infty$
we have
\begin{equation}
f(r) \sim p r^{-11/2} \cos (\alpha \log r) + q r^{-11/2} \sin (\alpha \log r),
\label{asymex}
\end{equation}
where $\alpha = \frac{1}{2} \sqrt{4 l^2 \lambda -9}$ is a constant. We
find that for large $r$, counting powers of $r$ gives:
\begin{equation}
h_{ab} h^{ab} \sqrt{g} \sim f^2 r^{10} \sim \frac{1}{r},
\end{equation}
so that $\langle h ,h \rangle$ does not converge. We define $f_n(r)$
by
\begin{equation}
f_n(r) = \rho_n(\log r) f(r).
\label{fn}
\end{equation}
Setting $h_{00}=f_n$ generates a perturbation which we call $h^n_{ab}$ via the constraint
equations and we claim that
\[
\frac{\norm{(\DL-\lambda I)h^n_{ab}}}{\norm{h^n_{ab}}} \to 0, \qquad
\mathrm{as} \quad n \to \infty.
\]
It is easily seen that $\norm{h^n_{ab}} \to \infty$ since the integral in
$\norm{h^n_{ab}}$ is bounded below by the integral for $\norm{h_{ab}}$
cut off at \mbox{$r=\rme^{n}$}. It suffices then to show that
$\norm{(\DL-\lambda I)h^n_{ab}}$ is bounded as $n \to
  \infty$. Clearly \mbox{$(\DL-\lambda I)h^n_{ab}=0$} on $\{r_+\leq r\leq
    \rme^n\} \cup \{\rme^{n+1}\leq r\}$ so we need to estimate
\begin{equation}
\int_{r=\rme^n}^{\rme^{n+1}}\left[\left(\DL-\lambda
  I\right)h^n\right]_{ab}\left[\left(\DL-\lambda
  I\right)h^n\right]^{ab} \rmd V = \norm{(\DL - \lambda I)h^n}^2,
\end{equation}
for large $n$. In this limit, we can use the large $n$ asymptotic
expansions to estimate the leading order behaviour of this
term. Substituting these in, we find that:
\begin{equation}
\norm{(\DL - \lambda I)h^n}^2 \sim K
\int_{\rme^n}^{\rme^{n+1}}(L\tilde{f}_n)^2 r^{10} \rmd r,
\label{asnm}
\end{equation}
with
\begin{equation}
L\tilde{f}_n = r^2\tilde{f}''_n + 12 r\tilde{f}'_n +(28+l^2\lambda)\tilde{f}_n,
\end{equation}
the leading order expansion of (\ref{diffeq}) and where $\tilde{f}_n$
is defined by:
\begin{equation}
\tilde{f}_n(r) =  (p r^{-11/2} \cos (\alpha \log r) + q r^{-11/2} \sin
(\alpha \log r))\rho_n(\log r).
\end{equation}
Under a change of variables $x=\log r$ the integral in (\ref{asnm})
becomes:
\begin{eqnarray}
\fl \int_{e^n}^{e^{n+1}}(L\tilde{f}_n)^2 r^{10} dr &=& \int_n^{n+1} \left\{
\rho_n''(x)(p \sin \alpha x + q \cos \alpha x) + 2 \alpha \rho_n'(x) (q \sin \alpha x - p \cos \alpha x)
\right\}^2 dx \nonumber \\ \fl
 &=& \int_0^{1} \left\{
\rho_0''(y)(p \sin \alpha (y+n) + q \cos \alpha (y+n))
\right.\nonumber \\ \fl &&
\hspace{0.5cm}\left. + 2 \alpha \rho_0'(y) (q \sin \alpha (y+n) - p \cos \alpha (y+n))
\right\} ^2 dy, \label{bded}
\end{eqnarray}
since $\rho_n(x)=\rho_0(x-n)$. Now for all $n$ the integrand is
bounded since
\begin{eqnarray*}
\abs{\rho_0'(x)} &\leq&6, \\
\abs{\rho_0''(x)} &\leq&60,
\end{eqnarray*}
thus there exist constants $C$ and $\tilde{n}$ such that
\begin{equation}
\norm{(\DL - \lambda I)h^n} \leq C, \qquad \forall \quad n \geq \tilde{n}.
\end{equation}
We can now apply lemma \ref{lem} to the operator $\DL$ and we find that for $\lambda > 9/4l^2$,
$(\DL - \lambda I)^{-1}$ is unbounded, hence \mbox{$\{\lambda > 9/4l^2\}
\subseteq C\sigma(\DL)$}.

This result makes no use of the value of $m$ or the location of the
zeroes of $A(r)$, so it holds for all three of the instantons
considered here. It is also possible to perform this analysis for the
other $SU(2)$ invariant modes of the perturbation and we find exactly
the same condition on $\lambda$ in order that it be in the continuous spectrum.

\section{Conclusions}

We have studied the semi-classical stability of the AdS-Taub-Nut
instantons and the two branch family of AdS-Taub-Bolt instantons and
we have found a negative mode of (\ref{licn}) for the \mbox{AdS-Taub-Bolt$^-$}
instanton, implying that this instanton is unstable. For the
AdS-Taub-Nut and AdS-Taub-Bolt$^+$ instanton we have argued that they
are (linearly at least) stable. We have also justified this by
considering a family of regular metrics, not necessarily satisfying
Einstein's equations, which contains these instantons as special
cases. This gives an intuitive sense of how the saddle-node
bifurcation in the $S-\mu$ plane arises, as well as the
Hawking-Page type bifurcation which occurs. We have also found that for all three instantons the continuous
spectrum includes the ray $\{\lambda > 9/4l^2\}$.

\ack

I would like to thank Gary Gibbons for proposing this project
and for useful discussions along the way. I am grateful to Sean
Hartnoll for advice and helpful comments. Many thanks to Gustav
Holzegel for much useful help and advice. I would also like to gratefully
acknowledge funding from PPARC.

\appendix
\section*{Appendix}
\setcounter{section}{1}
The following are a set of differential equations governing the
$J=M=K=0$ piece of the tensor harmonic decomposition of the equation
\[
 -\del{e} \nabla^e h_{ab} - 2 R_{acbd}h^{cd} = \lambda h_{ab},
\]
for the metric
\[
d s^2 = \frac{1}{A(r)} d r^2 + 4 N^2 A(r) \sigma_3^2 + B(r)
(\sigma_1^2 + \sigma_2^2).
\]
We suppress the $r$ dependence of $A(r)$ and $B(r)$ and use a prime to
denote differentiation with respect to $r$. These are taken from
\cite{Young:1984dn}, with some typos corrected.

{\bf 11 Component}
\begin{equation}
\label{11cpt}
\overline{a}(r) h''_{11} + \overline{b}(r) h'_{11} +
\overline{c}(r) h_{00} + \overline{d}(r) h_{11} + \overline{e}(r) h_{22}+ \overline{f}(r) h_{33}
= -\lambda h_{11},
\end{equation}
with
\begin{eqnarray*}
\overline{a}(r) &=& A, \\
\overline{b}(r) &=& A' - A \frac{B'}{B}, \\
\overline{c}(r) &=& -2 A (A' B') + \frac{4 A^2
  B'^2}{B} - 4A^2 B'', \\
\overline{d}(r) &=& \frac{2}{B} -\frac{A B''}{B} - \frac{A'
  B'}{B} + \frac{1}{2} A \left( \frac{B'}{B} \right)^2 -
\frac{4 N^2 A}{B^2} -\frac{1}{2N^2A}, \\
\overline{e}(r) &=& \frac{1}{2N^2A} - \frac{4N^2A}{B^2} - \frac{1}{2}
A \left( \frac{B'}{B} \right)^2, \\
\overline{f}(r) &=& - \frac{A'B'}{8 N^2 A} + \frac{1}{B}.
\end{eqnarray*}

{\bf 22 Component}
\begin{equation}
\label{22cpt}
a(r) h''_{22} + b(r)h'_{22} + c(r) h_{00} + d(r) h_{11} +
e(r) h_{22} + f(r) h_{33} = -\lambda h_{22},
\end{equation}
with
\[
\begin{array}{rclcrcl}
\displaystyle a(r) &=&\displaystyle \overline{a}(r), && \displaystyle
d(r) &=& \displaystyle \overline{e}(r), \\
\displaystyle b(r) &=&\displaystyle \overline{b}(r), &\hspace{1cm}
\mbox{and} \hspace{1cm}& \displaystyle
e(r) &=& \displaystyle \overline{d}(r), \\
\displaystyle c(r) &=&\displaystyle \overline{c}(r), && \displaystyle
f(r) &=& \displaystyle \overline{f}(r). \\
\end{array}
\]
Where the overbar denotes the coefficients for (\ref{11cpt}).

{\bf 00 Component}
\begin{equation}
\label{00cpt}
a(r) h''_{00} + b(r)h'_{00} + c(r) h_{00} + d(r) X +
f(r) h_{33} = -\lambda h_{00},
\end{equation}
with
\begin{eqnarray*}
a(r) &=& A, \\
b(r) &=& 3 A' + \frac{AB'}{B}, \\
c(r) &=& \frac{A'^2}{2A} + \frac{A'B'}{B} -  A \left(
\frac{B'}{B} \right)^2 + A'', \\
d(r) &=& - \frac{A'B'}{8AB^2} + \frac{B'^2}{4B^3} -
\frac{B''}{4B^2}, \\
f(r) &=& \frac{A'^2}{32N^2A^3} - \frac{A''}{16N^2A^2},
\end{eqnarray*}
and $X = h_{11}+h_{22}$.

{\bf 33 Component}
\begin{equation}
\label{33cpt}
a(r) h''_{33} + b(r)h'_{33} + c(r) h_{00} + d(r) X +
f(r) h_{33} = -\lambda h_{33},
\end{equation}
with
\begin{eqnarray*}
a(r) &=& A, \\
b(r) &=& -A' + \frac{AB'}{B}, \\
c(r) &=& 8N^2AA'^2 - 16N^2 A^2 A'', \\
d(r) &=& \frac{16N^4A^2}{B^3} - \frac{2N^2AA'B'}{B^2}, \\
f(r) &=& -A'' + \frac{A'^2}{2A} - \frac{A'B'}{B} - \frac{4N^2A}{B^2},
\end{eqnarray*}
and $X = h_{11}+h_{22}$.

{\bf Constraint Equations}

The above system of equations is consistent with the transverse and
tracefree conditions on $h_{ab}$. This gives us the constraint
equation
\begin{equation}
\label{tracefree}
A h_{00} + \frac{1}{4B} (h_{11} + h_{22}) + \frac{1}{16 N^2 A} h_{33}
= 0,
\end{equation}
coming from the traceless condition $h^a{}_a = 0$, and
\begin{equation}
\label{transverse}
A h'_{00} + \left ( \frac{3}{2}A' + \frac{A B'}{B}
\right) h_{00} - \frac{B'}{8{B^2}} \left( h_{11} + h_{22} \right) -
  \frac{A'}{32 N^2 A^2} h_{33} = 0,
\end{equation}
which comes from the transverse condition $\del{a}h^{ab}=0$.

\end{document}